\documentclass{article}

\usepackage{PRIMEarxiv}

\usepackage[utf8]{inputenc} 
\usepackage[T1]{fontenc}    
\usepackage{hyperref}       
\usepackage{url}            
\usepackage{booktabs}       
\usepackage{amsfonts}       
\usepackage{nicefrac}       
\usepackage{microtype}      
\usepackage{lipsum}
\usepackage{fancyhdr}       
\usepackage{graphicx}       
\graphicspath{{media/}}

\usepackage{algorithmic}
\usepackage{graphicx}
\usepackage{textcomp}
\usepackage{xcolor}
\usepackage{float}
\usepackage{multicol} 
\usepackage{subcaption}
\usepackage{hhline}
\usepackage{breakurl}
\usepackage{listings}

\usepackage{svg}
\svgpath{{./images/}}
\usepackage{caption}
\usepackage[utf8]{inputenc} 
\usepackage[T1]{fontenc}    
\usepackage{hyperref}       
\usepackage{booktabs}       
\usepackage{amsfonts}       
\usepackage{nicefrac}       
\usepackage{microtype}      
\usepackage{lipsum}
\usepackage{graphicx}
\graphicspath{ {./images/} }
\usepackage{amsmath}
\usepackage{multirow}
\usepackage{adjustbox}
\usepackage{array}
\usepackage{nicematrix}
\usepackage{makecell}

\usepackage{xurl}

\usepackage[section]{placeins}
\def\BibTeX{{\rm B\kern-.05em{\sc i\kern-.025em b}\kern-.08em
    T\kern-.1667em\lower.7ex\hbox{E}\kern-.125emX}} 
\title{\Large \bf Where the Polyglots Are:\\ How Polyglot Files Enable Cyber Attack Chains and Methods for Detection \& Disarmament}

\author{Luke Koch
\\kochlr@ornl.gov
\\Oak Ridge National Laboratory
\\Oak Ridge, TN, USA
\and Sean Oesch
\\oeschts@ornl.gov
\\Oak Ridge National Laboratory
\\Oak Ridge, TN, USA
\and Amul Chaulagain
\\chaulagaina@ornl.gov
\\Oak Ridge National Laboratory
\\Oak Ridge, TN, USA
\and Jared Dixon
\\dixonjm@ornl.gov
\\Oak Ridge National Laboratory
\\Oak Ridge, TN, USA
\and Matthew Dixon
\\dixsonmk@ornl.gov
\\Oak Ridge National Laboratory
\\Oak Ridge TN, USA
\and Mike Huettal
\\huettelmr@ornl.gov
\\Oak Ridge National Laboratory
\\Oak Ridge, TN, USA
\and Amir Sadovnik
\\sadovnika@ornl.gov
\\Oak Ridge National Laboratory
\\Oak Ridge, TN, USA
\and Cory Watson
\\watsoncl1@ornl.gov
\\Oak Ridge National Laboratory
\\Oak Ridge, TN, USA
\and Brian Weber
\\weberb@ornl.gov
\\Oak Ridge National Laboratory
\\Oak Ridge, TN, USA
\and Jacob Hartman
\\hartmanj@ainfosec.com
\\Assured Information Security
\\Rome, New York, USA
\and
Richard Patulski
\\patulskir@ainfosec.com
\\Assured Information Security
\\Rome, New York, USA}
\newcommand\blfootnote[1]{%
  \begingroup
  \renewcommand\thefootnote{}\footnote{#1}%
  \addtocounter{footnote}{-1}%
  \endgroup
}
\begin{document}
\maketitle
\blfootnote{Notice: This manuscript has been authored [or, co-authored] by UT-Battelle, LLC, under contract DE-AC05-00OR22725 with the US Department of Energy (DOE). The US government retains and the publisher, by accepting the article for publication, acknowledges that the US government retains a nonexclusive, paid-up, irrevocable, worldwide license to publish or reproduce the published form of this manuscript, or allow others to do so, for US government purposes. DOE will provide public access to these results of federally sponsored research in accordance with the DOE Public Access Plan (\url{http://energy.gov/downloads/doe-public-access-plan}).}

\begin{abstract}
A polyglot is a file that is valid in two or more formats. Polyglot files pose a problem for malware detection systems that route files
to format-specific detectors/signatures, as well as file upload and sanitization tools. In this work we found that existing file-format and embedded-file detection tools, even those developed specifically for polyglot files, fail to reliably detect polyglot files used in the wild, leaving organizations vulnerable to attack. To address this issue, we studied the use of polyglot files by malicious actors in the wild, finding $30$ polyglot samples and $15$ attack chains that leveraged polyglot files. In this report, we highlight two well-known APTs whose cyber attack chains relied on polyglot files to bypass detection mechanisms. Using knowledge from our survey of polyglot usage in the wild---the first of its kind---we created a novel data set based on adversary techniques. We then trained a machine learning detection solution, PolyConv, using this data set. PolyConv achieves a precision-recall area-under-curve score of $0.999$ with an F1 score of $99.20$\% for polyglot detection and $99.47$\% for file-format identification, significantly outperforming all other tools tested. We developed a content disarmament and reconstruction tool, \textit{ImSan}, that successfully sanitized $100$\% of the tested image-based polyglots, which were the most common type found via the survey. Our work provides concrete tools and suggestions to enable defenders to better defend themselves against polyglot files, as well as directions for future work to create more robust file specifications and methods of disarmament.

\end{abstract}

\section{Introduction}
\begin{figure}[h]
  \centering
  \includegraphics[width=.7\columnwidth]{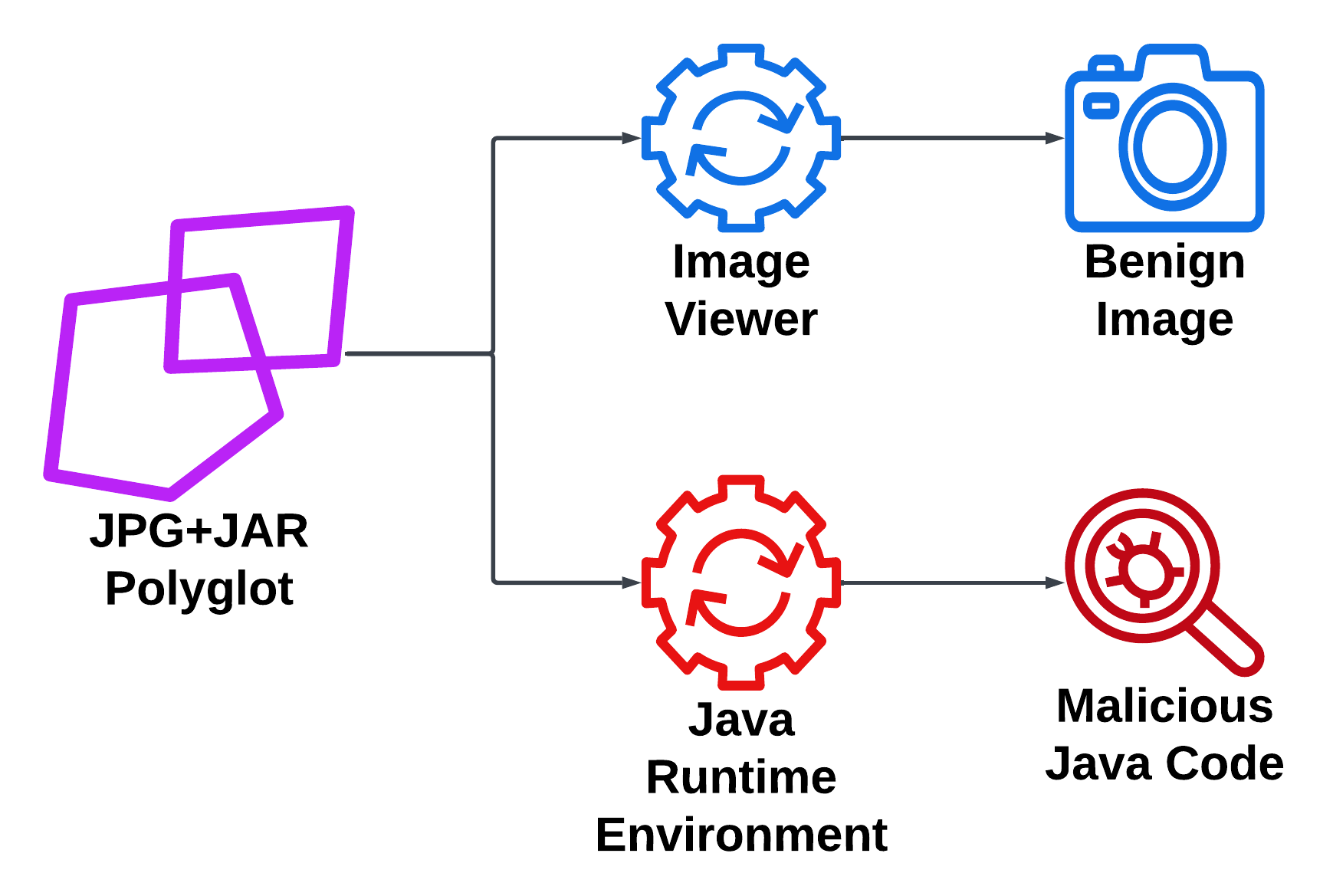}
  \caption{Functionality of a polyglot file is determined by the calling program, which can be explicitly provided or automatically determined by the operating system's auto-launch settings.}
  \label{Fig:function}
\end{figure}
\begin{figure}[h]
  \centering
\includegraphics[width=\columnwidth]{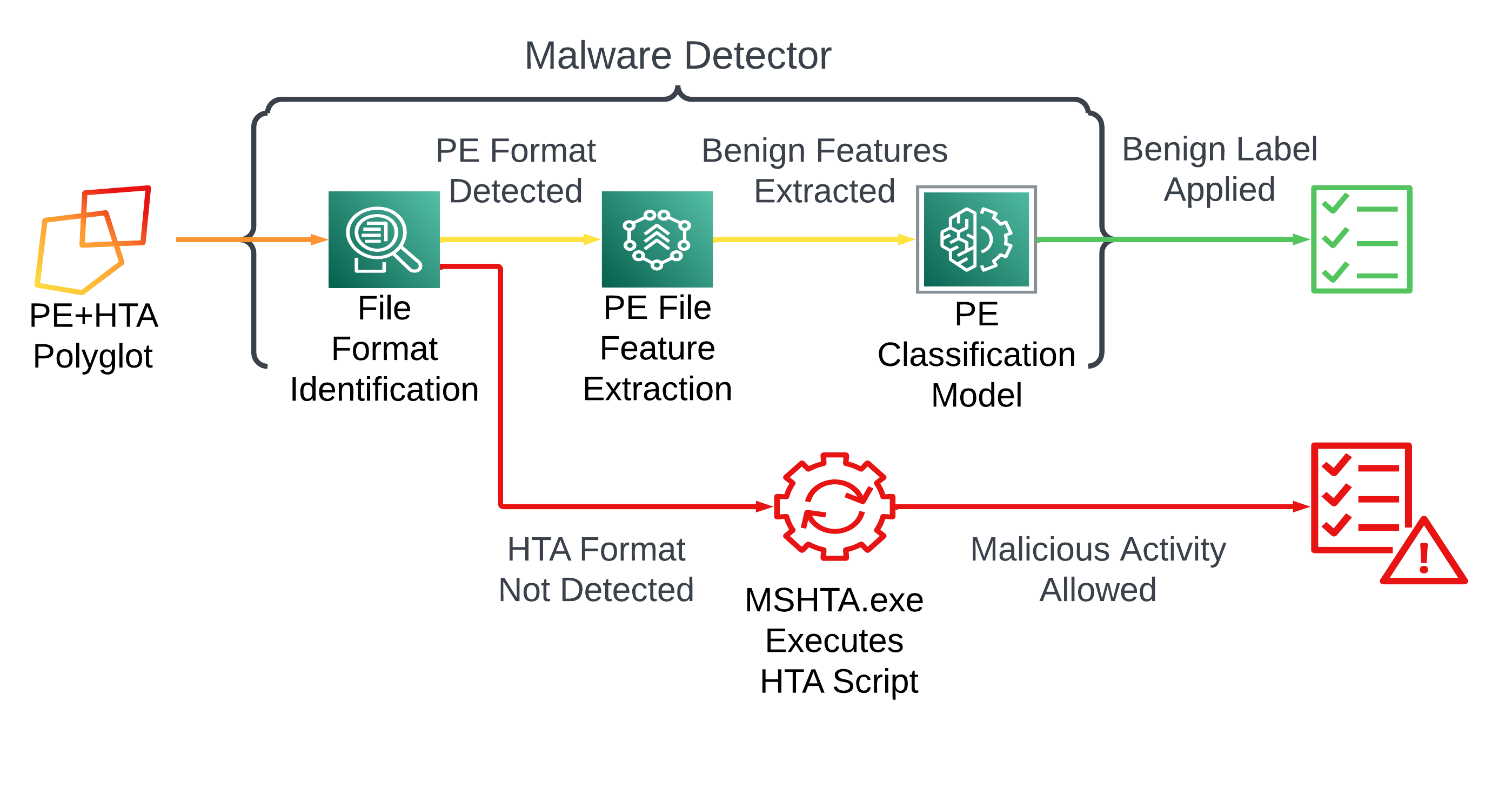}
  \caption{Since polyglot files simultaneously conform to multiple formats, they can evade correct format identification. This in turn allows them to evade format-specific feature extraction or signature matching, thereby evading malware detection. Therefore, some preprocessing should be done to either filter/quarantine polyglot files prior to feature extraction or route them to multiple format-specific malware detectors so all functional components of the polyglot are analyzed.}
  \label{Fig:threat}
\end{figure}

A polyglot file simultaneously conforms to two or more file-format specifications. This means the polyglot file can exhibit two completely different sets of behavior depending on the calling program, as depicted in Figure \ref{Fig:function}. This dual nature poses a threat to endpoint detection and response tools (EDR) and file-upload systems that rely on format identification prior to analysis. As shown in Figure~\ref{Fig:threat}, a polyglot can evade correct classification by first evading format identification. If only one format is detected, then the sample may not be routed to the correct feature-extraction routine (in the case of machine learning-based detectors) or compared to the correct subset of malware signatures (in the case of signature-based malware detection). 
As evidence that existing commercial off-the-shelf (COTS) endpoint detection and response tools are vulnerable to polyglots, we point to Bridges et al.~\cite{bridges2020beyond}, who demonstrated that 4 competitive COTS tools detected 0\% of the malicious polyglots in the test data. 

Standardized formats for files play a key role in cybersecurity. 
By first identifying the format of an unknown sample, they allow malware detection tools to extract the most discriminate and robust features from an unknown sample. This allows the detection tool to discard unimportant bytes that can be manipulated to alter classification in an adversarial attack~\cite{kolosnjaji2018adversarial, Demetrio_2021}. 
However, this feature-extraction process introduces a vulnerability; the correct format must be detected in order to route the file to the correct feature extractor. 
Even when a detector does not use machine learning and instead relies upon signatures for detection, the need to maintain a high throughput encourages EDR tools to only search for signatures that correspond to the detected format~\cite{abusing}.

As prior researchers~\cite{abusing, corkami_gtfo, clunie2019dual, ortiz2019hipaa, dicom-poly, popescu2012hiding} have demonstrated, polyglot files can be crafted that are fully valid (execute as intended) in multiple formats. 
To date, however, no comprehensive study of polyglot usage by malicious actors in the wild and/or methods of detecting said polyglots has been undertaken. 
In this paper, we set out to answer four key research questions related to polyglot usage and mitigation: 

\textbf{RQ1: } \textit{How are polyglots currently used by threat actors in the wild? This includes the role the polyglot fills, the formats of the donor files, and the combination method used to fuse the donors together.}

\textbf{RQ2: } \textit{Can we train a detector to effectively filter or reroute polyglots prior to ingestion by a malware detection system?}

\textbf{RQ3: } \textit{Does this detector outperform existing file type detection, file carving, and polyglot-aware analysis tools at detecting polyglot files?}

\textbf{RQ4: } \textit{Given the prevalence of image-based polyglots in adversary usage and the relative simplicity of image formats, what tools can we provide to defenders to address image-based polyglots in their existing workflows?}

To address \textbf{RQ1}, we reviewed open-source intelligence feeds (see Section \ref{survey_methods} for methods) that detail adversary tactics, techniques and procedures (TTP), finding that polyglots have played an important role in a number of malicious campaigns by well-known advanced persistent threat (APT) groups.
Polyglot files allowed the malicious actors to covertly execute malicious activity and extract sensitive data by masquerading as innocuous formats. 
In Section~\ref{section: wild}, we provide an overview of the different roles polyglots played in each campaign, detail the file combinations used, and provide a detailed description of several high profile examples.
To address \textbf{RQ2-RQ4}, we first created a tool, \textit{Fazah}, for generating polyglots that mimic the examples seen in the wild. Although there are other possible format combinations, our goal with this tool was to mimic, as closely as possible, the formats and combination methods used by real-world threat actors.
Using this tool, we then created a data set of polyglot and  normal (referred to hereafter as monoglot) files for  training and testing. See Section~\ref{section: dataset} for a full description of the data set. 

To address \textbf{RQ2}, we tested machine learning models to solve both the binary and the multi-label classification problems, achieving an F1 score of 99.20\% for binary classification and 99.47\% for multi-label classification with our deep learning model PolyConv. 
To address \textbf{RQ3}, we evaluated five commonly used format identification tools on this dataset: \textit{file} \cite{file}, \textit{binwalk} \cite{binwalk}, \textit{TrID}, \textit{polydet}, and \textit{polyfile}.  These tools were selected because of their use in existing cybersecurity tools or claim to detect polyglot files.
We evaluated the performance of these tools at both binary  and multi-label classification. 
In our context, binary classification determines whether a file is a polyglot or a monoglot. Multi-label classification, on the other hand, identifies all formats to which the file conforms. 
We found that existing tools did not exceed an F1 score of 93.32\% at binary classification and 83.74\% at multi-label classification.

See Section~\ref{section: mlforpoly} for details regarding our ML based approaches and Section~\ref{section: comparison} for a comparison of ML-based approaches to existing file-format identification tools. 

As detailed in Section~\ref{sub: imagebased}, to address \textbf{RQ4} we developed and tested a CDR tool for sanitizing image-based polyglots since these were the most common vector for polyglot malware. We also tested YARA rules for detecting extraneous content in image files.
We found that the YARA rule approach did not generalize well to all formats that can be combined with an image, especially the more flexible scripting formats like Powershell or JavaScript. However, they may be use in high-throughput use cases where deploying a deep learning model is not feasible.
A more effective approach is to strip all extraneous content from images using a content disarmament and reconstruction (CDR) tool. Our CDR tool, \textit{ImSan}, was able to sanitize all of the image polyglots in a random subset of our image polyglots. A subset was used so we could manually verify the results.

The following provides a summary of our contributions:
    \begin{itemize}
        \item \textbf{RQ1:} The first, to our knowledge, survey of polyglot usage by malicious actors in the wild, demonstrating that polyglot files are an actively used TTP by well-known malicious actors. 
        Utilizing the results of this study, we created a tool, \textit{Fazah}, to generate polyglots using formats and combination methods exploited by malware authors in the wild.
        We then used \textit{Fazah} to generate a dataset of polyglots and monoglots to evaluate existing detection methods and train polyglot detection models. 
  
        \item  \textbf{RQ2:} Utilizing this novel dataset, we trained a deep learning model, PolyConv, that can distinguish between polyglots and monoglots with an AUC score over 0.999. 
        We also created a multi-label model that reports all of the detected formats in monoglot and polyglot files, enabling analysts to quickly determine the nature of a threat or route the suspicious file to multiple format-specific detection systems. 
        \item \textbf{RQ3:} We provide a comparison of our polyglot detection models with existing file-format identification and carving tools, some of which are polyglot aware.
        This evaluation shows that existing methods for detecting file type manipulation are inadequate and often fail to detect polyglot files, even with special flags set that are meant to ensure multiple file types are detected.
        
        \item \textbf{RQ4:} For image-based polyglots, which are common in the wild, we explored YARA rules and content disarmament and reconstruct (CDR) tools, finding that our \textit{ImSan} CDR tool was 100\% effective while the YARA rules did not compete with our deep learning detector. They may, however, be of use in high throughput situations.
        
    \end{itemize}

\section{Related Work}

\subsection{Polyglot Detection}
Bridges et al. conducted an in-depth evaluation of four leading COTS tools~\cite{bridges2020beyond}. Among the test data were 199 malicious JPG+JAR polyglots that went completely undetected by all 4 tools. While we can not prove why these tools failed across the board, we can surmise---based on the unusual 0\% detection rate---that the failure occurred in the file-type identification that must occur prior to feature extraction. If the files were interpreted as JPG (the benign component) rather than JAR (the malicious component), it is unlikely that the malicious JAR content was analyzed. This provides a plausible explanation for the complete detection failure. Therefore, to solve the problem of malicious polyglot detection, the problem of correct file type identification should first be solved.

The machine learning models FiFTy and Sceadan---a support vector machine (SVM) and a deep learning model, respectively---were released by researchers for file-format identification \cite{fifty, sceadan}. However, neither tool was designed with polyglots in mind or trained on a dataset containing them. 

\subsection{Polyglot Creation}\label{creation}
Jana and Shmatikov demonstrated a number of attacks that exploited discrepancies in file type inference  and file parsing. Specifically, they found that polyglots---referred to as "ambiguous files conforming to multiple formats" \cite{abusing}---evaded detection by 20 out of 36 malware detectors. Using the open-source ClamAV tool as an example, they point out that malware detectors may terminate format inference at the first match, extracting features and/or checking malware signatures only for the first format detected in the polyglot. Jana and Shmatikov argue that exhaustively testing incoming files against all possible formats would introduce an unacceptable overhead. Flexibility in format specification and parser tolerance of malformed files are also presented as reasons why simply improving existing tools is difficult.

Ange Albertini demonstrated that a wide variety of files can be combined into polyglots~\cite{corkami_gtfo}. He created an open-source tool known as \textit{mitra} that can create 4 types of polyglot from a wide range of files. He defined the four types thusly:
\begin{itemize}
    \item \textit{Stack}: File B is appended to the end of file A
    \item \textit{Parasite}: File B is placed inside comment markers of file A
    \item \textit{Zipper}: Both files are placed within one another's comment markers
    \item \textit{Cavity}: File B is placed inside a padding area of file A
\end{itemize}

\subsection{Polyglot Exploitation}
A number of previous academic works demonstrated the risk polyglots may pose. For example, a DICOM file is an image archive format designed for medical use. The format was designed to be flexible so medical staff could combine a variety of image formats into a single file for a patient~\cite{dicom-intent}. However, that flexibility means a DICOM file will tolerate combination with a Windows Portable Executable (PE) file to create a malicious polyglot~\cite{ortiz2019hipaa}. This polyglot could allow an adversary to propagate their malicious PE through a medical network, activating the PE component through a second stage of the attack. 

In an attack on data integrity, Popescu demonstrated that a PDF+TIFF polyglot can bypass digital certification verification~\cite{popescu}. In this scenario, the attacker sends a valid request (PDF file) for a bank transfer to a target. The attacker's goal is to change the amount authorized in the PDF without invalidating the certification applied by the target. When the victim opens the file, auto-launch settings intepret the file as a PDF and present the legitimate PDF contents to the victim. 

The victim then applies a digital signature that protects the file contents from any future change, and returns the file to the attacker. However, the file is also a TIFF file. The TIFF is an image of the same PDF, albeit with a much larger money transfer authorized. The attacker does not edit the contents of the file (which would break the signature). They merely change the file extension, switching the auto-launch behavior from opening the PDF contents to opening the TIFF contents, before sending the file on to a hypothetical bank. 

Since the file contents have not changed, the digital certification is still valid. When the bank opens the file, auto-launch behavior shows the larger fraudulent TIFF transaction rather than the proper PDF amount the victim agreed to when they signed the file.


\section{\textbf{RQ1:} Polyglot Exploitation in the Wild}\label{section: wild}
Thanks to Bridges et al. \cite{bridges2020beyond}, we know polyglots can evade detection by COTS tools. However, the extent to which malicious actors employ polyglots has never, to our knowledge, been published before. Do malicious actors use polyglots in their attack chains? What role do polyglots play within an attack chain? What file formats and combination methods were utilized in these attacks?
To address these questions we conducted a survey of threat intelligence feeds, collecting file hashes of polyglot samples and information on the roles played by these files within attack chains.
For the file hashes and a list of the sources used in this survey, see Table~\ref{tab:survey_hashes} and Table~\ref{tab:survey_sources} in Section~\ref{appendix}.

\subsection{Survey Methods}\label{survey_methods}
The survey, performed between November 2022 and January 2023, focused on identifying the role of a polyglot file within a threat actor's cyber-attack chain. 
We used publicly available independent sources, general search engines and threat intelligence feeds (e.g., ORKL, X) to gather a wide range of information security reports and articles. Those sources were searched using the following terms: \textit{polyglot}, \textit{combined}, and \textit{contained}. We found that the term \textit{polyglot} is not always  utilized in reports. We therefore had to manually distinguish between reports of true polyglots (two or more valid formats in one file) and other forms of digital steganography. A number of reports described malware that contained a valid format along with an oft-encrypted set of malicious instructions. We do not consider these files as polyglots because the malicious instruction can only be correctly interpreted when passed as input to another component of the malware rather than a parser conforming to a published standard. 

For each true polyglot found, we used our knowledge of threat operations to determine the role the polyglot played in the cyber-attack chain. Lastly, the online malware databases, VirusTotal and MalwareBazaar, were used to obtain the actual polyglot samples whenever hashes of the polyglot were provided in a report. The file hashes and sources from our survey of open-source intelligence can be found in the appendix in Tables~\ref{tab:survey_hashes} and ~\ref{tab:survey_sources}, respectively. 

\subsection{Role of Polyglot Files in Cyber Attack Chains} 

The survey discovered fifteen examples of a threat actor using a polyglot file in their cyber-attack chain, along with 30 distinct polyglot files. 
According to MITRE's Adversarial Tactics, Techniques, and Common Knowledge (ATT\&CK) framework, polyglots are primarily utilized for \textit{Defense Evasion} (MITRE ATT\&CK TA0005). Polyglot files also fall under the \textit{Obfuscated Files or Information} (MITRE ATT\&CK T1027) heading since these files conceal hidden functionality by appearing to conform to only one file format. We obtained 30 polyglot samples from VirusTotal and MalwareBazaar using the file hashes specified in the reports.

For the purpose of establishing a formal taxonomy for polyglot files, we refer to polyglots as having an overt format and a covert format. The overt format is the format the file presents as (e.g., matches the extension) while the covert format is not apparent without analysis. In most cases, a polyglot consists of a malicious file combined with a benign one; however, in some cases we found that both file formats play a role in advancing the malicious attack chain, as in the HTA+CHM polyglot utilized by IcedID in Section~\ref{iced}. Therefore, we instead refer to polyglots as combining an overt format with a covert format. A summary of the found-in-the-wild samples is provided in Table~\ref{tab:combos}. In Appendix \ref{formats} we discuss the capabilities of interest that each file format provides to the malware author (camouflage, non-standard execution path, etc.) to understand why these combinations exist in the wild and how they fill a desired role in  attack chains.




We selected two cyber attack chains to demonstrate how well-known APTs utilize polyglots to reach the next step in their cyber attack chains. A third attack chain can be found in Appendix \ref{chain}. CVE numbers and MITRE ATT\&CK references are provided where applicable.
\subsubsection{IcedID} 
\label{iced}
IcedID is a banking trojan that, according to Check Point's Global Threat Index, was the fourth most widespread malware variant in 2022~\cite{ice}. The trojan uses an evolving variety of methods to establish initial access. One of these methods relies on a polyglot formed by combining a CHM and an HTA file.

\begin{figure}[h] 
  \centering 
  \includegraphics[width=.8\columnwidth]{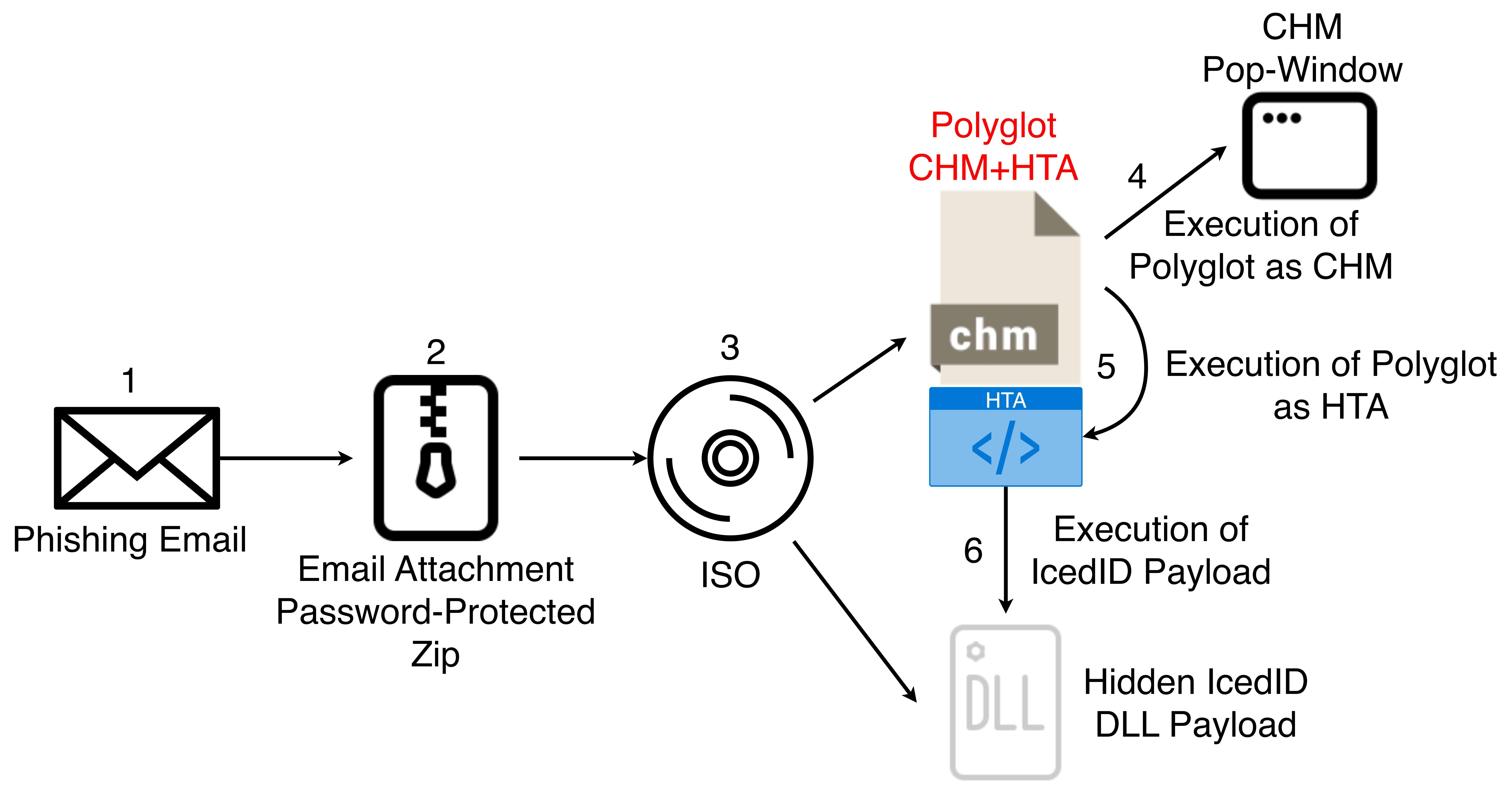} 
  \captionof{figure}{IcedID Attack Chain}\par\medskip 
  \label{Fig:icedid} 
\end{figure}

The attack chain is illustrated in Figure~\ref{Fig:icedid}. It begins with a password-protected Zip file attached to a phishing email. The Zip contains an ISO file which exploits  CVE-2022-41091 to evade flagging by Microsoft's \textit{alternate data stream} (ADS) defensive mechanism~\cite{motw}. 

The ISO file in turn contains two files: a DLL (hidden by default on Windows) and a CHM+HTA polyglot. The polyglot masquerades as a CHM file which presents a benign decoy window when executed. The Microsoft compiled HTML (CHM) format used for software documentation. Each file consists of a number of HTML pages organized into a document that is compressed into a binary stream. As with any HTML page, CHM files may download/execute other files or run Powershell/Javascript commands when viewed. 

In the background, this CHM file starts a MSHTA.exe process with itself as the input. This new process executes the malicious component of the polyglot, the HTA file, which in turn  launches the hidden DLL file that contains the actual IcedID payload. 

\subsubsection{Andariel/Lazarus} 
\begin{figure}[h] 
  \centering 
  \includegraphics[width=.8\columnwidth]{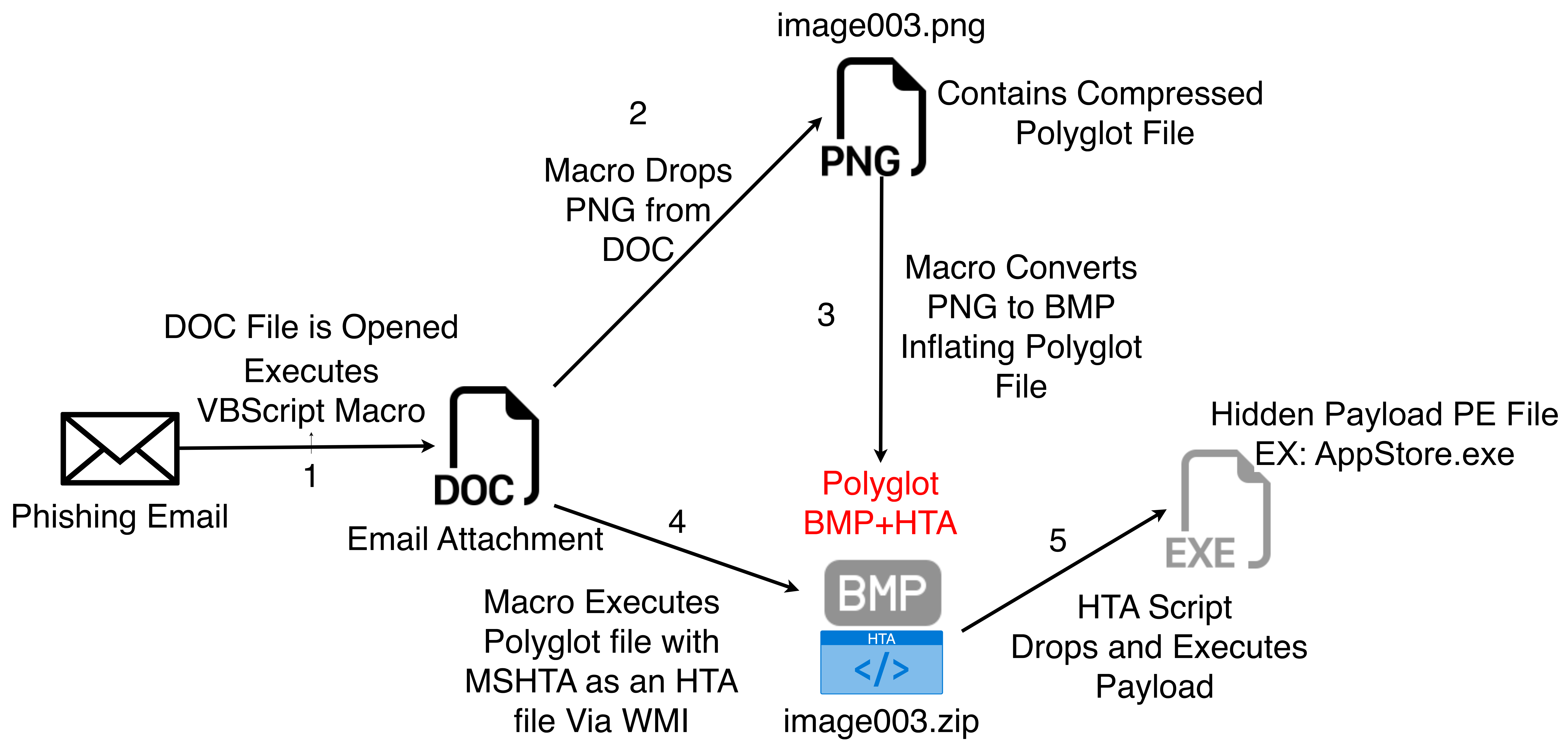} 
  \captionof{figure}{Andariel/Lazarus Attack Chain}\par\medskip 
  \label{Fig:andariel} 
\end{figure}

Lazarus (of which Andariel is a subgroup) is an advanced threat group that has operated out of North Korean since 2009~\cite{mitre}. In 2021 attack chains connected to this group utilized polyglots to infect systems with a Remote Access Trojan (RAT)~\cite{malwarebytes, kaspersky}; this process is illustrated in Figure~\ref{Fig:andariel}.

This attack chain typically begins with a phishing email that has an attached malicious Microsoft Word Document (DOC) file (MITRE ATT\&CK T1566). When the DOC file is launched, a macro begins execution (MITRE ATT\&CK T1204.002). First, the macro drops a PNG file to the \textit{Temp} directory. The image data in the PNG file is a compressed polyglot file.

Next, the DOC macro converts the PNG file to a BMP file, which has the intended side effect of decompressing the contents (MITRE ATT\&CK T1140). The DOC Macro does this by leveraging the Windows Image Acquisition (WIA) Automation Layer Objects: ImageFile and ImageProcess~\cite{imagefile, imageprocess}.

After conversion, the DOC Macro saves the BMP as a zip file by giving it a zip extension. However, the file is actually a BMP+HTA polyglot, with the HTA covert contents appended to the end of the overt BMP data. Finally, the DOC Macro executes the polyglot file as an HTA file using the MSHTA application via the Windows Management Instrumentation (WMI) Service (MITRE ATT\&CK T1059, T1047).

WMI is used so that the resulting process does not appear to be a child of the DOC process. The HTA file drops its payload, a hidden PE file, into a hidden folder. Finally, the HTA file launches the PE file which provides a foothold on the target system for future exploitation.      

    \section{Wild Polyglots: A Polyglot Data Set Based on Malicious Usage in the Wild}\label{section: dataset}
This section describes how we created our data set based on our survey of polyglot usage in the wild (\textbf{RQ1}) using the \textit{Fazah} tool in order to address \textbf{RQ2-RQ4}. 

\subsection{Fazah: A Polyglot Generation Framework}

Having uncovered which formats have been used in real-world malicious polyglots, we created a data set consisting of monoglot and polyglot files conforming to these formats. Our first step was to create a framework for generating polyglots by combining donor files. Our goal for this tool was to mimic format and combination methods found in the wild rather than demonstrate all possible combinations. 
\begin{table}[h]
\caption{Polyglot Formats Deployed Maliciously in the Wild}
\centering
\label{tab:combos}
\begin{tabular}{|l|l|}
\hline
Covert Format & Overt Format\\
\hhline{|=|=|}
HTA & \makecell[l]{JPEG, PNG, BMP, GIF, LNK,\\ PE, MSI, RAR, Zip, TTF,\\ RAR, CHM, PDF} \\ 
\hline
PHP & \makecell[l]{JPEG, PNG, BMP, GIF, TTF,\\ RAR, Zip, LNK, PDF} \\
\hline
PHAR & JPEG, PNG, BMP, GIF \\
\hline
JavaScript & GIF, BMP \\
\hline
PowerShell & JPEG, BMP, GIF \\
\hline
Zip & JPEG, PNG, GIF, PDF \\
\hline
JAR & \makecell[l]{JPEG, PNG, GIF, PDF,\\ MSI} \\
\hline
RAR & JPEG, PNG, BMP, GIF \\
\hline
BMP & Zip, JAR\\
\hline
\end{tabular}
\end{table}
The \textit{Fazah} framework is a modular tool written in Python that can currently generate $46$ format combinations using $8$ \textit{covert} formats. The combination method---\textit{stack} and a variety of \textit{parasites}---is derived from reports of malicious use in our survey and varies between \textit{covert} format. As discussed in the survey, malicious actors use polyglots either to disguise malicious content using a less suspicious format (images) or add hidden functionality (scripts). Since image formats typically use comment markers, \textit{parasites} are commonly used by malicious actors. \textit{Stacks}, meanwhile, are the simplest and easiest method for malicious actors to implement, working well with script and archive formats. Files with distinct comment markers (necessary for \textit{zippers}) are quite rare. Of the common (but by no means exhaustive) set of formats we tested, only DCM combined with either PDF/GIF/ISO could result in a \textit{zipper}. Similarly, we found that only ISO paired with PE/PNG/GIF yielded \textit{cavities}. This does not preclude their use in malicious campaigns, but places them beyond scope for our goal of emulating known attack chains. Table~\ref{tab:combos} provides the format pairings that \textit{Fazah} can turn into polyglots.
Given the possibility for malicious abuse of the framework, \textit{Fazah} will not be published publicly at this time. 

\subsubsection{Wild Polyglots Data Set Creation and Contents}
\begin{figure}[h]
  \centering
\includegraphics[width=\columnwidth]{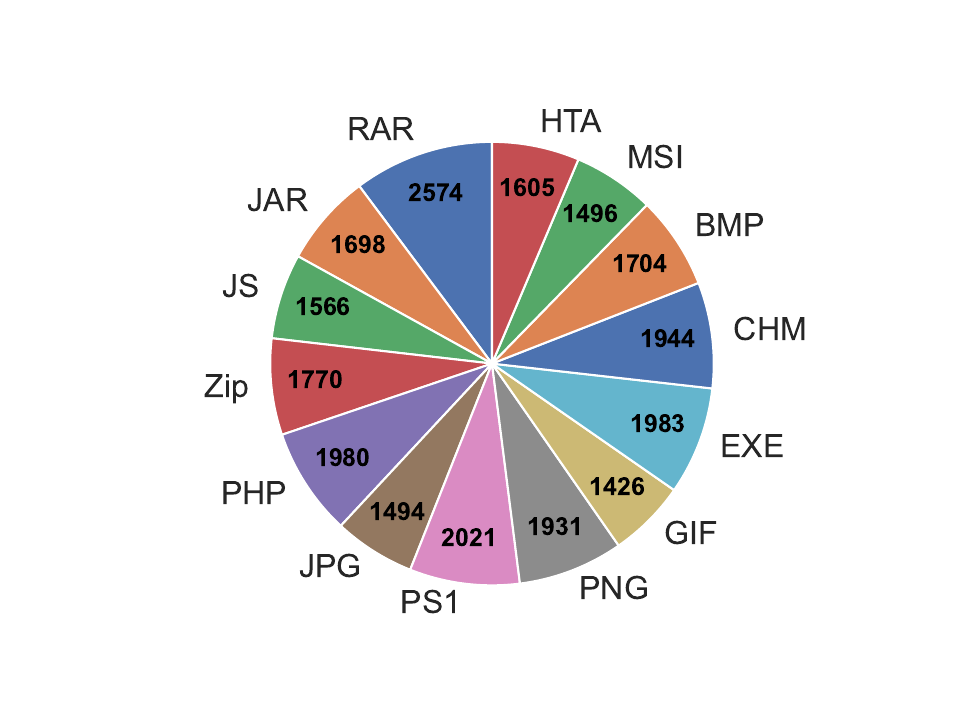}\vspace{-30pt}
  \caption{File counts for the monoglot  formats in the Wild Polyglots training data. }  
  \label{Fig:mono}

\end{figure}

\begin{figure}[h]
  \centering
\includegraphics[width=\columnwidth]{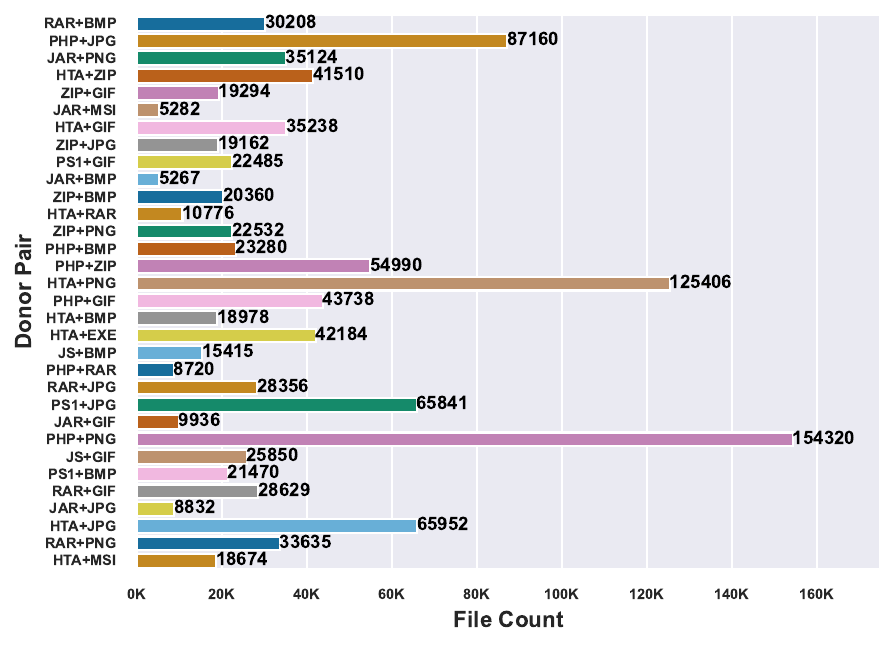}
  \caption{File counts for each of the 32 polyglot combinations in the Wild Polyglots training data.}
  \label{Fig:poly}
\end{figure}

We collected benign files conforming to 13 common formats using Github's search API: BMP, EXE, GIF, HTA, JAR, JPG, JS, MSI, PHP, PNG, PS1, RAR, ZIP. Using a held-out set of donor files, we created 32 types of polyglots organized according to which 2 types of donor files were combined to create the polyglot file. We kept all donor files separate from the train and test set to ensure that the models did not cheat by learning that data added to a monoglot in the training set is a polyglot. Table~\ref{tab:traintest} provides an overview of the Wild Polyglots data set. Figures~\ref{Fig:mono} and~\ref{Fig:poly} breakdown the formats contained in the monoglot and polyglot training sets, respectively.  
Since our objective was to train a polyglot detector rather than a malware detector, we only utilized benign files. We first scanned the files we scraped for malware and removed any suspicious samples. Next, we removed any scraped files whose extension did not match the file contents (e.g., a JPEG with a .png extension) or if the file could not be parsed by an appropriate utility (e.g., Pillow for images). We erred on the side of inclusion for highly flexible scripting language formats like HTA. Since MSHTA.exe is tolerant of a high degree of malformation, we felt it unwise to exclude malformed HTA from our training data. 
\begin{table}[]
\centering
\caption{Wild Polyglots Data Set Contents}
\label{tab:traintest}
\begin{tabular}{|l|l|l|}
\hline
      & Train & Test \\   \hhline{|=|=|=|}
Monoglot & 25192    &  9975 \\ \hline
Polyglot  & 1148604     & 213109   \\ \hline

\end{tabular}
\end{table}


\section{\textbf{RQ2:} Using Machine Learning for Polyglot Detection}\label{section: mlforpoly}
This section explores using machine learning to detect polyglot files. Section~\ref{development} chronicles our development process as we tested different ML model architectures and experimented with improvements to the feature space. Section~\ref{testing} presents the results from out best-performing models compared to existing tools. 
\subsection{Ml-based Detection Development}\label{development}
Our first objective was to determine which machine learning architecture and feature set were most effective at detecting polyglots. Toward this end, we created a small ($\sim70,000$ files) initial data set using the \textit{mitra} tool (described in Section~\ref{creation}) prior to the development of our \textit{Fazah} tool. On this preliminary data set, we tested a Support Vector Machine, Random Forest, GradBoost, CatBoost, LightGBM, and MalConv. With the exception of MalConv \cite{malconv}, these models used the byte histogram as their only feature. The byte histogram is a vector of length 256 where the value stored at each index corresponds to the number of times that byte value occurs in the input file. This feature vector is agnostic with respect to file formats since all digitally stored files are a string of bytes. We found that, on this preliminary data set, MalConv and CatBoost were the top performers. 

We focused further development on MalConv and CatBoost, labeling our improved versions PolyConv and PolyCat, respectively. At this point, we trained and tested both models on our survey-informed Wild Polyglots data set; all results and figures reported in this paper refer to the Wild Polyglots data set. We found that, for PolyCat but not PolyConv, adding the mime-type output of the \textit{file} utility improved results. Although \textit{file} was not competitive at detecting polyglots (see Section~\ref{tool_methods}) or at identifying both formats contained within, it was extremely accurate at identifying the first  format contained in the file. Therefore, we augmented PolyCat's feature space with a $1$-hot encoding of the mime-type output from \textit{file}. We found further improvement by adding the $8000$ most common bigrams and trigrams extracted from each file using an overlapping window. Thus, the final feature space for PolyCat consisted of the byte histogram, the 1-hot encoding of the mime-type from \textit{file}, and the most common bigrams and trigrams. 

MalConv is an oft-cited deep learning classifier designed to detect malware \cite{malconv}. We trained the model from scratch to identify polyglots rather than to identify malware. None of the polyglots in our data set were malicious in order to guarantee that the model learned to detect multiple formats rather than malicious content.  Since the model is trained on raw bytes rather than format-specific features (e.g., the EMBER feature set for PE files~\cite{ember}), MalConv's architecture is well-suited to the polyglot detection problem which requires a format-agnostic approach. In lieu of a fixed feature-extraction routine, the model takes in raw bytes and learns an encoding (first layer) as well as a set of filters (the convolution layers) to recognize significant byte patterns. MalConv also features an attention and gating mechanism intended to filter out extraneous information in the raw bytes. 

\begin{figure}[h]
  \centering
\includegraphics[width=\columnwidth]{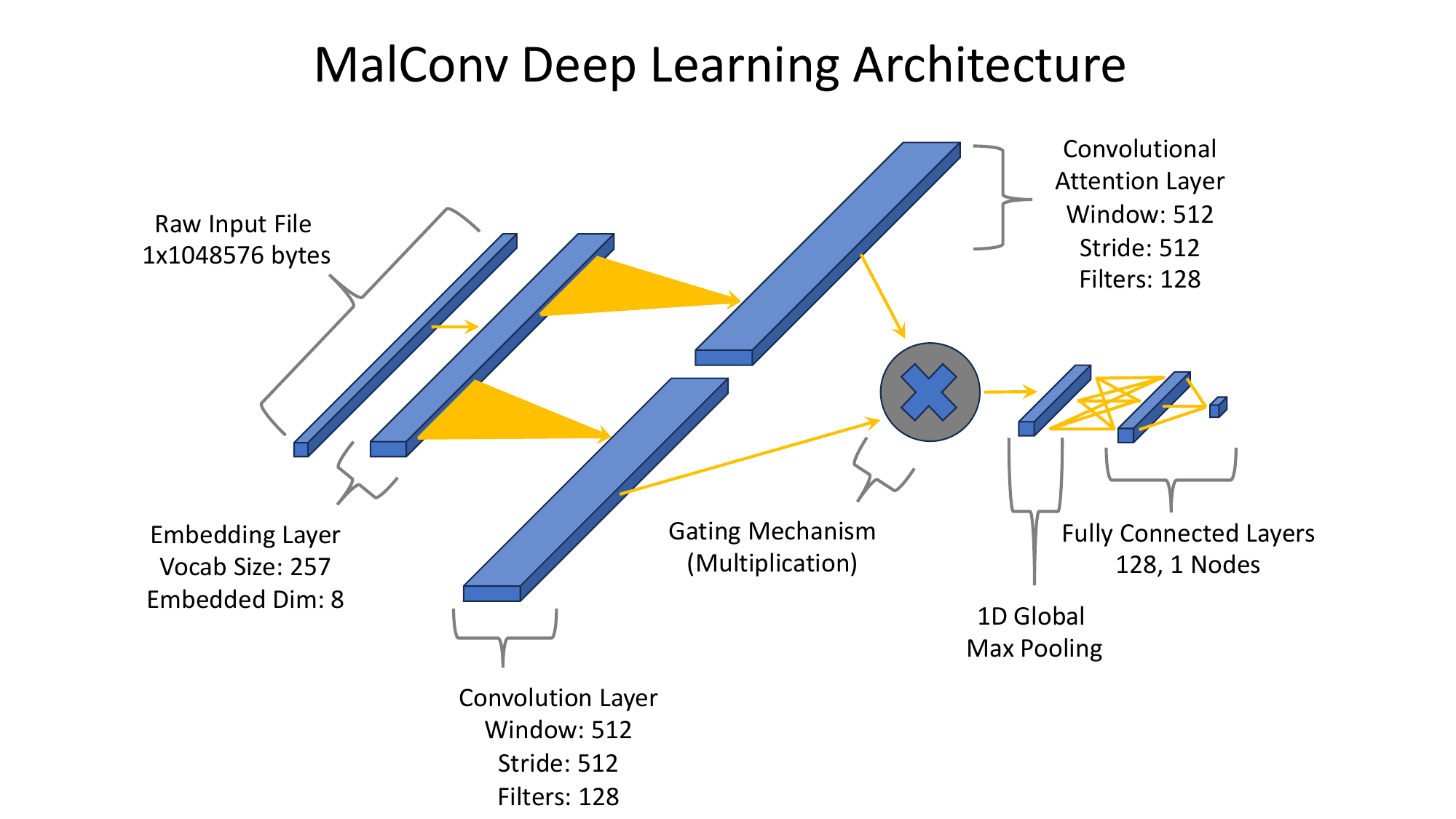}
  \caption{MalConv Architecture}
  \label{Fig:malconv_diagram}
\end{figure}

\begin{figure}[h]
  \centering
\includegraphics[width=.9\columnwidth]{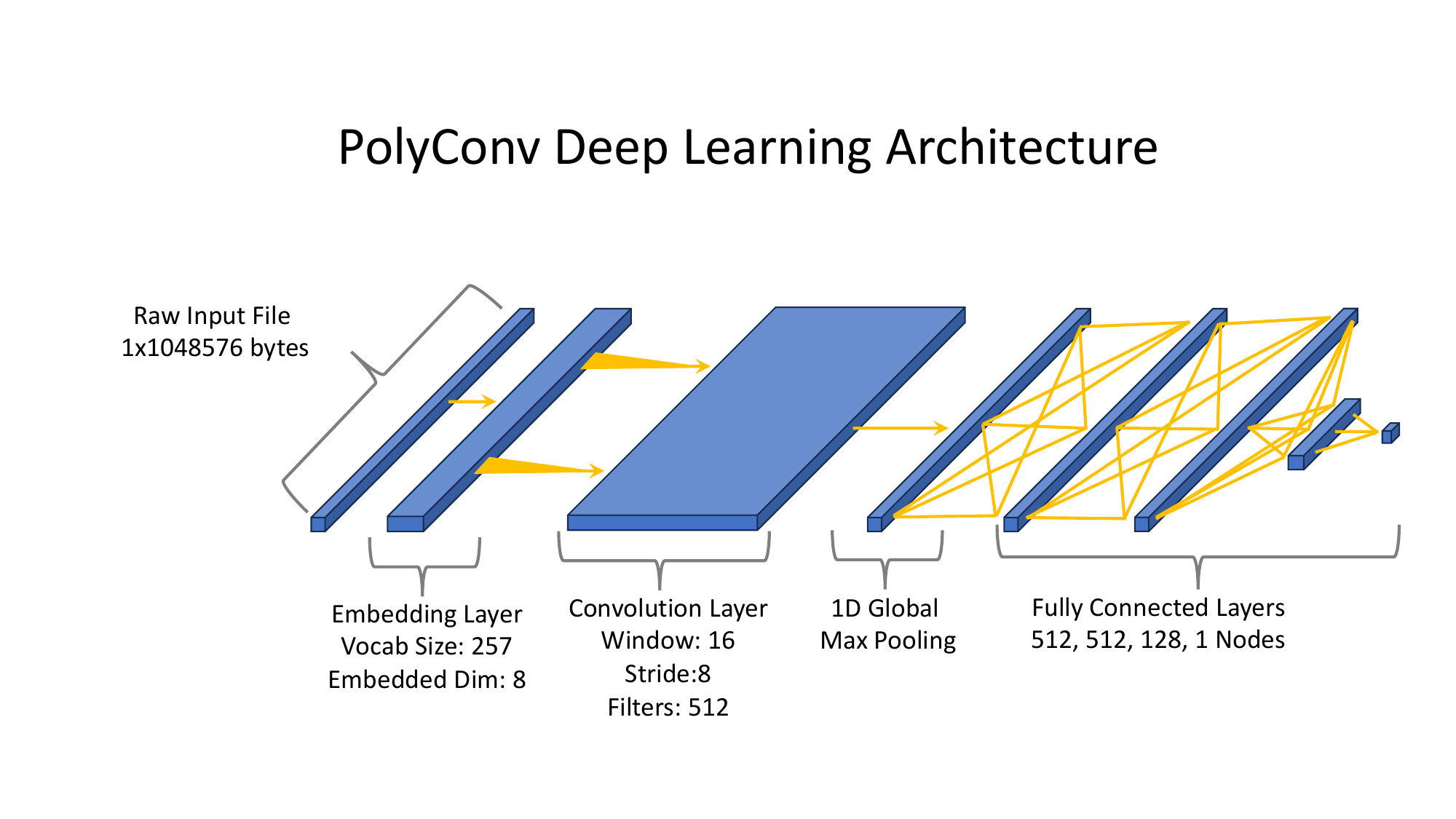}
  \caption{PolyConv Architecture}
  \label{Fig:polyconv_diagram}
\end{figure}

We experimented with changes to the architecture in order to make it more effective at our novel task, yielding the PolyConv model mentioned above. The original architecture of MalConv is presented in Figure~\ref{Fig:malconv_diagram} while PolyConv's architecture is presented in Figure~\ref{Fig:polyconv_diagram}. 

The changes we made to MalConv consist of the following:
\begin{itemize}
    \setlength{\itemsep}{1pt}
    \item Decreasing the window and stride from 512 bytes to 16 and 8 bytes, respectively, in order to capture the byte patterns of very short (in terms of bytes) script files hidden within larger files
    \item Removing the attention and gating mechanism as they did not seem to improve the results on our task
    \item Increasing the number of kernels in the remaining convolution layer to 512 in order to learn enough byte patterns to distinguish the wide variety of distinct formats upon which we trained the model
    \item Increasing the number of fully connected layers to 3 as a result of experimenting with different layers counts
    \item Increasing the number of nodes in each fully connected layer to 512, 512, and 128 as a result of experimenting with different node configurations
\end{itemize}
\subsection{Comparing ML-based Polyglot Detection Approaches}\label{testing}
\begin{figure}[h]
  \centering
\includegraphics[width=.7\columnwidth]{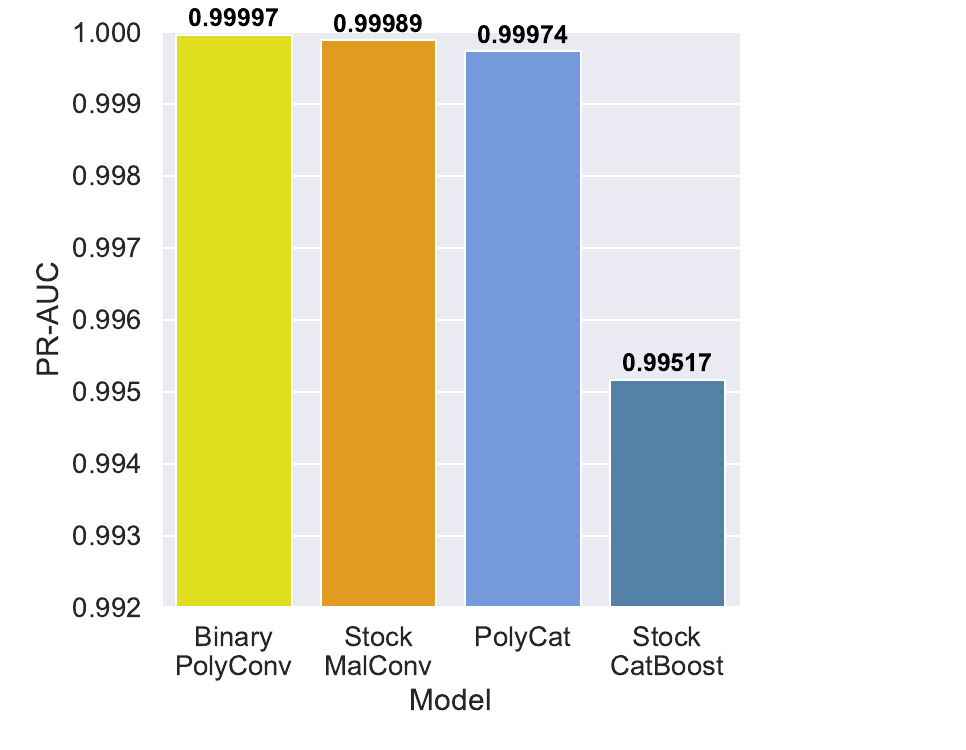}
  \caption{Precision-Recall AUC Scores: Our deep learning model, PolyConv, slightly outperformed the stock version of MalConv upon which it is based as well as  CatBoost and Polycat.}
  \label{Fig:prauc}
\end{figure}

We trained and tested PolyConv, MalConv, PolyCat, and CatBoost on our Wild Polyglots data set. For this comparison, we evaluated binary label (polyglot or monoglot) versions of the models. Since our data set is imbalanced, we used the precision-recall curve rather than the ROC curve to score our models. Therefore, our top model is the one with the highest PR-AUC on the Wild Polyglots test set.  PolyConv scored a PR-AUC of 0.99998, the highest score for all the models we evaluated. MalConv---when trained on this novel task---scored a slightly lower PR-AUC of 0.99989, outperforming both PolyCat and CatBoost. 
The model results are summarized in Figure~\ref{Fig:prauc}.
\section{\textbf{RQ3:} Comparison to Existing Signature-based File-format Identification Tools}\label{section: comparison} This section compares our best-performing polyglot detection model, PolyConv, to existing tools for format identification to determine which approach is best suited to identifying polyglot files and labeling their contents correctly.

Within the context of cybersecurity, there are two complimentary questions of paramount importance: detection and analysis. We trained two versions of our best-performing model, PolyConv, that differ only in the final layer to suit detection and analysis needs.

The first version is a binary classifier (polyglot or monoglot) for use in filtering out polyglots on an endpoint. This is intended for file upload services that only want to allow uploads of known formats, e.g., images. 

The second version is a multi-label classifier to identify all of the formats detected within a file. This provides two benefits. First, the labels can be used to route files to all applicable file-format feature extraction  or signature-matching routines rather than a single format-specific model or signature subset. This means that the remainder of an existing EDR tool's extraction and detection routines do not need to be altered. Second, the labels provide an analyst with introspection, revealing not only that the file is a polyglot but also which format-specific tools/routines they should use to examine the covert format(s) hidden in the polyglot. This is intended to reduce the response time necessary for secure operation center (SOC) analysts that must handle a high volume of alerts.
\subsection{Tools Tested}
\label{tool_methods}
We established a baseline for performance by testing existing file-format identification tools on the Wild Polyglots data set: \textit{file} \cite{file}, \textit{binwalk} \cite{binwalk}, \textit{TrID} \cite{trid}, and \textit{polydet} \cite{polydet}. We also evaluated \textit{polyfile} \cite{polyfile}, a DARPA-funded tool developed by Trail of Bits for detecting unusual files. 
Of the aforementioned tools, \textit{file} and \textit{TrID} are well-established signature-based utilities for file-format identification. VirusTotal, a widely used anti-virus aggregator (\url{www.virustotal.com}), utilizes \textit{TrID} when reporting detected formats. \textit{Binwalk} is a file-carving tool that has been used by analysts to find and extract hidden files. We selected these tool to establish a baseline because of their wide-spread use (\textit{file}), cybersecurity application (\textit{binwalk},\textit{TrID}), and polyglot-awareness (\textit{polyfile}, \textit{polydet}). We leave as future work a comparison to \textit{FiFTy} \cite{fifty} and \textit{Sceadan} \cite{sceadan}, as these detectors do not appear to be polyglot-aware, but might be re-trained in order to properly label polyglot files. Since \textit{file} outputs labels and not probabilities, the precision-recall curve is not an appropriate metric when comparing our deep learning model to existing tools. Instead, we used the F1 score that balances recall and precision. For any cybersecurity system deployable in the real-world, the ability to detect malware/polyglots (recall) must be tempered by a low probability of false positives (precision) to prevent red-flag fatigue. Therefore, we use F1 to provide a balanced evaluation.  \subsubsection{Binary Comparison}
\begin{figure}[h]
  \centering
\includegraphics[width=.7\columnwidth]{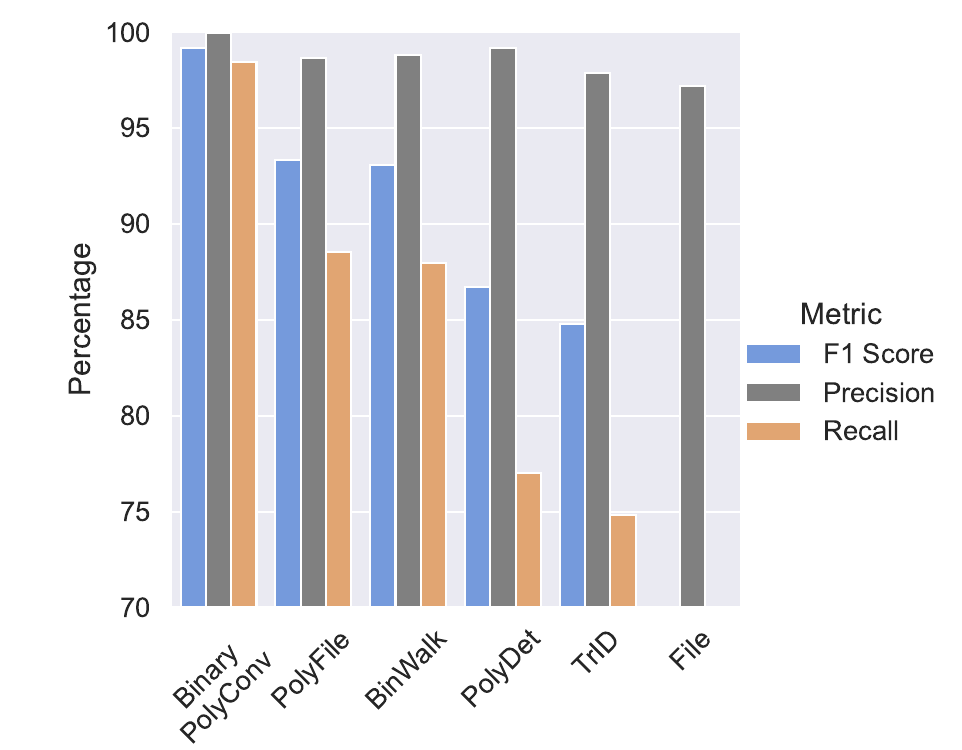}
  \caption{Binary Performance vs Existing Tools: PolyConv exceeded the F1 score, precision, and recall of all existing tools by a large margin.}

 \label{Fig:binary} 
\end{figure}

Figure~\ref{Fig:binary} considers the performance of each tool in a binary context, determining if the tool detects the presence of two or more formats in one file. \textit{TrID} aggressively speculates as to which formats are present in a file, assigning a percent score to each possibility. We therefore omitted the performance of \textit{TrID} as a multi-label detector as this behavior put it at a disadvantage compared to the other tools. 
As Figure~\ref{Fig:binary} demonstrates, none of the existing tools approached the F1 score, precision, or recall of our PolyConv deep learning model. All of the tools had a relatively high precision and low recall, indicating that false negatives were the primary cause of the low F1 scores. 

The recall for \textit{file} was lower than expected as the tool reported multiple formats when examining BMP, EXE, HTA, and PHP monoglots. The EXE false positives may have been caused by the presence of other files embedded as resources. Although it was outperformed by our PolyConv model, \textit{polyfile} was the best binary performer among the existing set of tools by F1 score. 

\subsubsection{Multi-label Comparison}
\begin{figure}[h]
  \centering
\includegraphics[width=.7\columnwidth]{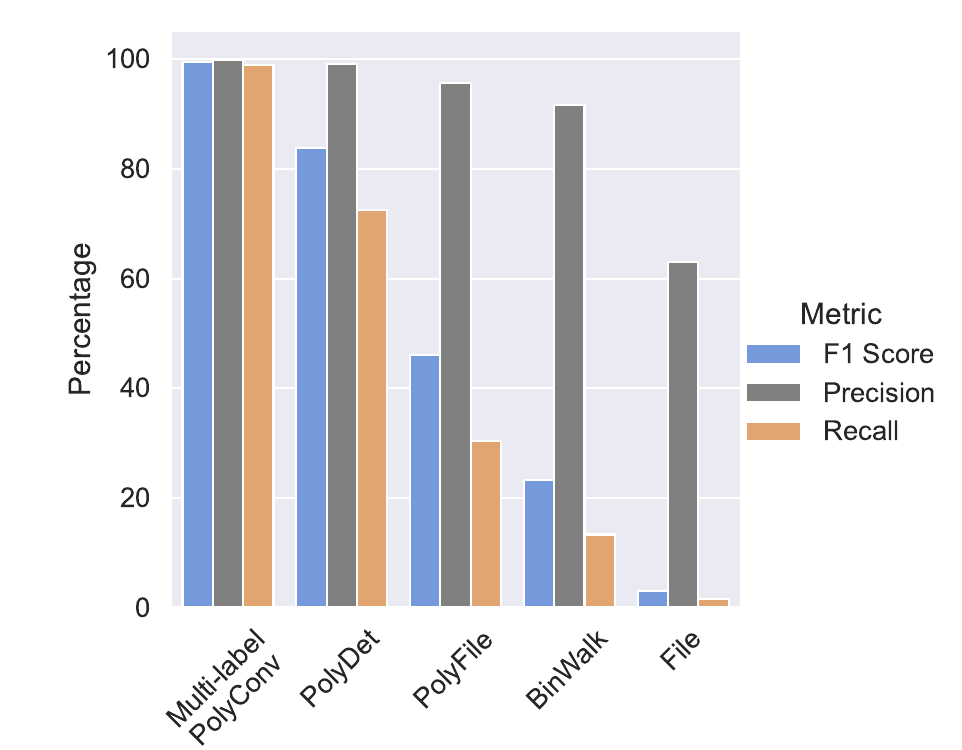}
  \caption{Multi-label Performance vs Existing Tools: PolyConv also proved more adept at correctly identifying all of the  formats contained within a file. Of the existing tools, \textit{polydet} provided the most reliable file-format identification.}
  \label{Fig:multi}
\end{figure}

Figure~\ref{Fig:multi} considers the performance of each tool in a multi-label context where a true positive means the tool correctly identified both the count and the exact formats present in each file. None of the tools performed well compared to the multi-label version of PolyConv. 

Of the existing tools, \textit{polydet} outperformed the other tools in all three metrics by a noticeable margin. With regard to the remaining tools, \textit{file}'s precision is unusually low given its widespread use and long development history. Upon examination, we found that \textit{file} did not differentiate between PowerShell and JavaScript files; instead, it applied the generic label of ASCII or Unicode text. This behavior almost exclusively accounted for the lower precision. 

The lack of required signatures for script files makes signature-based detection difficult for these script formats. Upon further inspection we found that \textit{polyfile} and \textit{polydet} share \textit{file}'s dependence on Libmagic, which labels PowerShell and JavaScript as either ASCII or Unicode text. While it might seem unfair to expect Libmagic to differentiate between different forms of ASCII or Unicode text, we consider it important for analysts to be aware of this opaque label. A harmless log file of unstructured ASCII text presents a very different level of danger compared to a functional JavaScript file. 
\section{\textbf{RQ4:} Methods for Addressing Image-based Polyglots}\label{sub: imagebased}
Given the prevalence of image-based polyglots in adversary usage and the relative simplicity of image formats, we developed tools for detecting and remediating polyglots that employ an overt image format.

We first tested YARA rules in the hopes that the comment markers/delimiters present in image files would allow for rule-based detection of extraneous content. However, we found that their recall of 82.08\% and F1 score of 90.15\% were too low to be useful except in situations where high throughput is tantamount. We then turned to the content disarmament and reconstruction approach. 
\subsection{\textit{ImSan}, a Content Disarmament and Reconstruction Tool for Image-based Polyglots}\label{imsan}
Content disarmament and reconstruction (CDR) tools present an alternative approach to the pre-processing filtering approach for which we have provided solutions. CDR tools allow an end user to strip all but the most trustworthy content from certain formats. Where highly flexible formats, like PDF, have proliferated, these tools have emerged to provide secure use of files that abuse the format flexibility.

Although we have not exhaustively examined this approach, we have developed an image sanitization tool to demonstrate the potential of CDR in disarming polyglots. 
Our tool, \textit{ImSan}, disarms image-based polyglots by stripping away all file contents that are not required to display the image. The process is quite straightforward:
\begin{enumerate}
    \item The image file is loaded into Pillow, a fork of the Python Imaging Library
    \item The image contents are then written to a new file with the option to strip all metadata activated
    \item The new image file has no extraneous content before/after the image contents (\textit{stack}/\textit{cavity} polyglot) or inserted into comment areas (\textit{parasite}/\textit{zipper} polyglot)
\end{enumerate}

\textit{ImSan} can disarm any of the formats that are fully supported (read/write) by Pillow: BLP, BMP, DDS, DIB, EPS, GIF, ICNS, ICO, IM, JPEG, JPEG 2000, MSP, PCX, PNG, PPM, SGI,SPIDER, TGA, TIFF, WebP, XBM. 
Note, \textit{ImSan} should be run in an isolated environment to ensure that no vulnerability in Pillow (2 CVE's reported in 2022) could allow a malicious image to gain execution when the image is parsed.

ImSan disarmed 100\% of the image polyglots in a subset (n=392) of image polyglots drawn randomly from the benign Wild Polyglots data set. 
A small subset was chosen so we could manually verify disarmament through visual inspection of the image's code rather than relying on one of our detectors. 
An evaluation of commercial CDR tools against polyglots (including those that are not image based) and the potential methods of circumventing CDR solutions, while out of scope for this work, would be a valuable direction for future work to explore. 
\section{Discussion}\label{discussion}
\subsection{Contribution Summary}
We presented the first, to our knowledge, survey of polyglot usage by malicious actors in the wild, demonstrating that polyglot files are an actively used TTP by well-known malicious actors, answering \textbf{RQ1}.

In order to answer \textbf{RQ2-RQ4}, we created a novel data set of polyglot and monoglot files based on file formats and file-combination methods utilized by malicious actors in the wild.

Using this Wild Polyglots data set, we evaluated a number of different machine learning models before focusing on the top two performers, PolyConv and PolyCat. we improved these two models via alterations to their architecture and feature space, respectively. 

We found that PolyConv, in both binary and multi-label versions, was effective at detecting polyglots and correctly labeling their contents \textbf{RQ2}, providing analysts with a tool to detect, reroute, and investigate potential polyglots.

PolyConv only slightly outperformed the model upon which it was based, MalConv, demonstrating that MalConv effectively learned to distinguish between polyglot and monoglot files when trained on this objective, despite being designed to detect malware. This is a novel use of MalConv considering that the model was designed to detect PE malware.    

Based on our experiments, the improvement from MalConv to PolyConv was due to the reduction of the window and stride size as well as increasing the number of filters and layers. We theorize that the much smaller window/stride allowed the model to learn filters that register even small areas of code with a distinct byte pattern. The need for more filters may be due to the wide variety of formats, each with their own distribution of unique byte patterns, upon which we trained.  On the other hand, removing the attention and gating mechanism did not reduce the model's classification performance. 

We answered \textbf{RQ3} by demonstrating that existing tools do not reliably detect polyglot files, even when designed with an awareness of polyglot files. 

To answer \textbf{RQ4}, we produced a set of YARA rules for detecting extraneous content in image files, but found their performance lacking. The rules are available upon request. We then created \textit{ImSan}, a content disarmament and reconstruction tool that sanitizes image files, demonstrating that it disarmed all of the image-based polyglots with which we tested it. 


\subsection{Limitations}
We cannot guarantee that our deep learning models will perform well on polyglots formed from file formats not included in the training data. 
File formats based on Open XML (Microsoft Office) are a common malware vector that have not yet been thoroughly explored as polyglot components. 
PolyConv is format agnostic so we hope that this model's release will prompt further research using additional file formats.

File size is a limiting factor for malware classification models trained on raw bytes rather than extracted features. Detection could be evaded by utilizing a polyglot whose overt format exceeds the full input capacity of the model, meaning the covert format would not be ingested. 
We tested head and tail scanning on our data set, but found that this did not improve results since the vast majority of our data set is within the maximum capacity of our PolyConv model. 
Head and tail scanning could still be evaded if an adversary inserted the second file near the middle of a particularly large first file or appended a large amount of data after the second file contents.

We also tested the YARA rule approach, but found it A) limited by the need to write novel rules for each possible  combination of file formats and B) the lack of required signatures in many flexible file formats. 

\subsection{Future Work}
Given the ubiquity of signature-based tools, the development and demonstration of a set of signatures or rules that  detect a very high percentage of functional script files as well as the ability to differentiate between different scripting languages would be valuable.  While JavaScript (JS) lacks a strict signature, it might be possible to craft a set of rules, perhaps using a tool such as AutoYARA~\cite{autoyara}, that detects all JS files that have enough control flow instructions to be functional/malicious. This would require the collection of a large number of files for each language, along with some level of assurance that the data set is extremely diverse with respect to the file contents/functionality. If Libmagic were able to reliably distinguish JavaScript and Powershell files from other forms of text, the performance of \textit{polyfile} and \textit{polydet} could improve markedly. We are not confident the rule-based approach is tractable, yet cannot rule it out. 

Since PolyConv utilizes a global max pooling layer, it is translation invariant. That said, a demonstration of its ability to generalize to novel insertion areas remains future work.  We consider translation invariance an important feature in order to future-proof a polyglot detector. Given the flexibility in file formats, it is possible that novel polyglot creation methods will emerge in the future that hide the second file in a novel area of the first file. Therefore, a demonstration that PolyConv is resilient in the face of novel combination methods would demonstrate that future models for polyglot detection should also be translation invariant. 

Future work should include the implementation of an intelligent method for subselecting or compressing large input files so they fit within the maximum capacity of a model trained on raw bytes. Head and tail scanning would catch data appended to the very end of the file, but could be evaded by inserting data earlier in the file or appending more benign content after the additional malicious content. Therefore, a more robust input reduction method should not follow a fixed pattern such as always scanning N bytes from the head and M bytes from the tail. Such a method may exist in other domains; we look forward to developments in this area. 

Finally, PolyConv needs to be trained and tested on the same wide variety of files as the ubiquitous \textit{file} utility in order to see widespread adoption. 
\subsection{Recommendations}
\subsubsection{Polyglot Detection}
Deep learning clearly outperformed other approaches to polyglot detection. Therefore, EDR tools or file upload services concerned with handling polyglots should invest in deep learning to identify and analyze incoming files. Our work demonstrates that the approach is effective and tractable. CDR tools should be used to strip extraneous content from files without loss of (benign) functionality, assuming said tool is fully compatible with the overt formats of incoming files. 

CDR tools should become ubiquitous for the formats where which which they are compatible. As for whether filtration or disarmament is the better approach, we would argue that defense-in-depth is the best approach. Enterprises should use commercial or open-source CDR tools on all file formats that said tools can successfully disarm and reconstruct reliably. For formats that are more complex or difficult to reconstruct reliably, enterprises should train PolyConv to catch any polyglots exploiting these formats. We suggest it is more tractable to train a detector to recognize a wide variety of byte patterns than it is to code reconstruction routines for all of the necessary file formats. 
\subsubsection{Existing Signature-based Tools}
Until deep learning models trained on the huge variety of files necessary for enterprise-scale use are available, we recommend utilizing \textit{polyfile} to filter suspected polyglots. For routing polyglots to multiple models or gaining introspection for analysis, we recommend using \textit{polydet}. Although neither of these tools were nearly as effective as our PolyConv model, they were the top performers among existing tools. Users should bare in mind that they do not discriminate between JavaScript, PowerShell, and plain text.

\subsubsection{File-format Specifications}
\label{specs}
File-format specifications and their practical implementations determine whether or not a polyglot can be created from two or more file formats. One way to continue to use signature-based detection would be to require all file formats to use a signature at the same fixed offset. Since two unique signatures cannot exist at the same offset, any combinations of file formats in a single file would not be valid/functional.

This solution do not appear practical as it would break backward compatibility for an enormous variety of files, e.g., Zip and DICOM formats which are intentionally flexible in this respect. The difficulty in securing flexible formats while preserving backward compatibility with existing files led to the development of CDR tools in the first place. 

Emerging file-format specifications, however, can benefit from our lessons learned. The chief difficulty lies in preventing a flexible format from being inserted into or appended to the novel file format. Toward preventing this, we recommend the following specifications as a starting point:
\begin{itemize}
    \item Digitally certify all files upon creation
    \item Require that officially supported parsing programs verify this certificate
    \item Ensure the certification area does not contain slack space that is ignored during the hash calculation
    \item Specify a file-format signature that must be present at offset zero
    \item Do not provide optional/slack spaces where extraneous data could be hidden
    \item Any "reserved for future use" fields should be validated to ensure they do not contain unexpected values
    \item Validate that any padding areas (due to memory paging) do not contain unexpected values
    
\end{itemize}

\section*{Acknowledgements}
This work was funded by a key intelligence community partner's Laboratory for Advanced Cybersecurity Research under a Memorandum of Agreement.

\bibliographystyle{ACM-Reference-Format}

\onecolumn
\clearpage
\section{Appendix}\label{appendix}
\subsection{Batloader/Zloader Cyber Attack Chain}\label{chain}

\begin{figure}[h] 
  \centering 
  \includegraphics[width=.8\columnwidth]{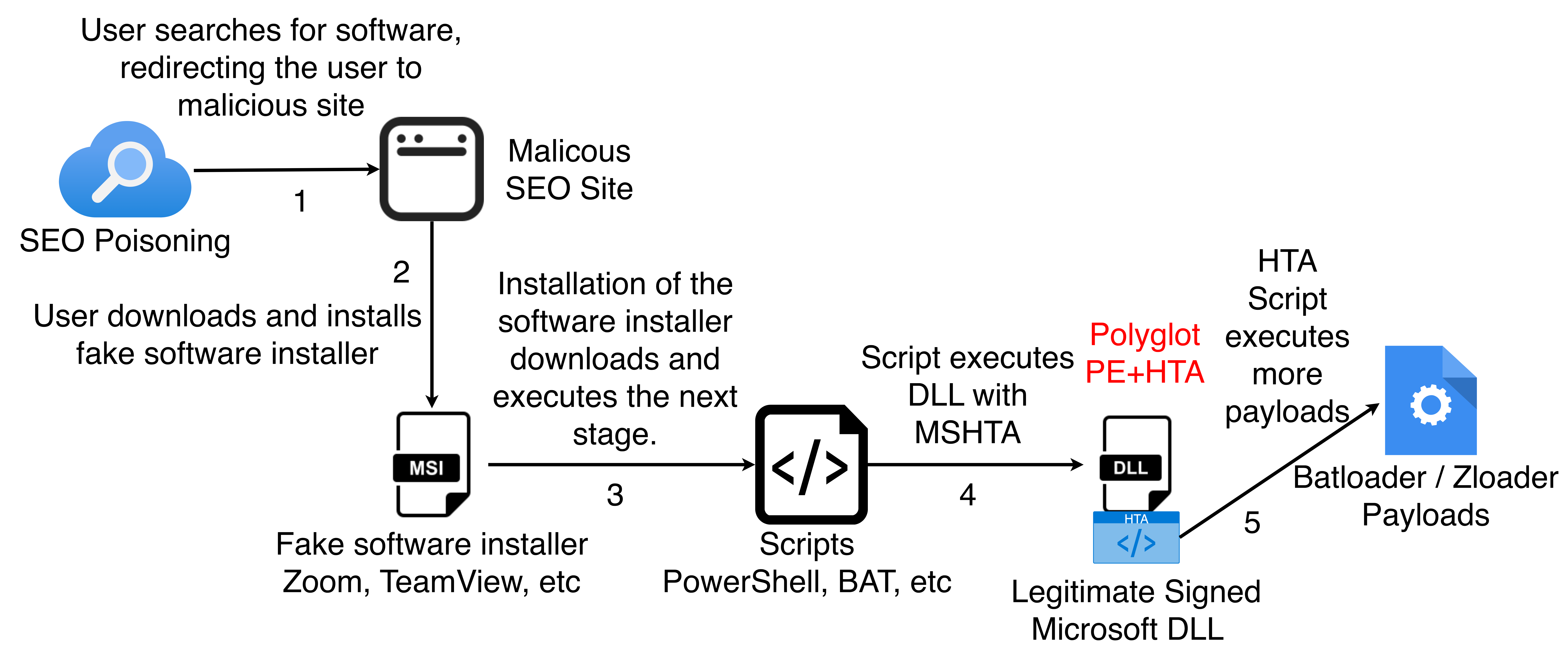} 
  \captionof{figure}{Batloader/Zloader Attack Chain}\par\medskip 
  \label{Fig:batloader} 
\end{figure}

Batloader and Zloader are two very similar pieces of malware that are used to gain initial access~\cite{mandiant, checkpoint-zloader}. The full attack chain is presented in Figure~\ref{Fig:batloader}; however, our discussion will focus on the role of the polyglot within that chain. This polyglot is formed by combining an HTA file with a Windows PE file. 

Windows PE files are the default executable for the Windows ecosystem. Since their format specification requires the bytes "MZ" to be present at offset zero, this format must be the first---by offset---ingredient in a polyglot in order to preserve functionality. PE polyglots can be created via the \textit{cave} or \textit{stack} method. The \textit{cave} method places the second file in a slack region of the PE. Candidate locations include the DOS Stub, after the last section table entry, or in the padding space after each section assuming the chosen region is large enough to contain the second file. The \textit{stack} method simply appends the second file to the end (also referred to as the overlay) of the PE file. 

In this particular example, an HTA file is added to the signature section of the PE file. Rather ironically, CVE-2020-1599 allows malware authors to add contents to the signature section without invalidating the signature since the contents of this area need to be writable in order to store the calculated signature. 

By preserving the validity of the rest of the file, the infected PE is able to operate with a higher degree of trust than it would otherwise~\cite{diaz2022monitoring}. Note that, although Microsoft addressed this vulnerability by creating an option to disallow extraneous data in the signature section, this option is turned off by default.  The higher degree of trust accorded to the signed PE allows the covert HTA to execute the final payloads in the Batloader/Zloader attack chain. 

\subsection{How File Formats Enable Polyglot Capabilities}\label{formats}

The following sections detail how each overt format was used in combination with a covert format to surreptitiously execute or stage a  malicious payload. This is followed by details on the roles filled by polyglot files in notable cyber-attack chains.  

\subsubsection{HTA} 
\label{wild_hta}
HTML Application (HTA) support in Windows is intended to make the Internet Explorer browser a Windows desktop development platform. It gives developers the flexibility to create full-scale applications using web-based technologies, such as HTML, JavaScript, and Visual Basic Script (VBScript) without following the strict security model of the browser.  

HTA contents are not executed directly; rather, they are fed to MSHTA.exe, a trusted signed Microsoft binary packaged with the Windows ecosystem, which then executes the contents. A few features of HTAs have attracted threat actors to include HTA files within their cyber-attacks.  

\begin{enumerate} 
  \item HTA files are loaded by a trusted, Microsoft-signed application, allowing attackers to bypass restrictions on application execution. 
  \item HTA files can be loaded remotely, allowing malicious activities to be run without even being copied to the target's disk.  
  \item MSHTA.exe has a generous parser, and does not require the HTA file signature \textit{<hta:application} for execution. MSHTA.exe will attempt to execute any HTML or VBScript/JavaScript code passed to the binary. This can make reliable identification of HTA files or fragments challenging for signature-based tools.  
\end{enumerate} 

The extensive usage of HTA files by attackers has led to an arms race between new detection methods and evasion techniques, with polyglots being one of the latest developments. A simple HTA attack can be detected using a rule which checks whether a MSHTA.exe process has been launched with an HTA file as input. This detection can be evaded by renaming/moving MSHTA.exe to a new name/location and by turning the HTA file into a polyglot that masquerades as a different file format. Since MSHTA.exe skips all data it does not understand, HTA files can be combined with a wide variety of file formats for obfuscation.

\subsubsection{PHP and PHAR} 

PHP is a popular programming language for web applications that provides dynamic rendering of web pages, database access, and many other features. The PHAR file format is the archive format of the language, comparable to JAR files within the Java ecosystem. As with HTA files, PHP has a generous parser that ignores a wide variety of syntax errors and invalid characters. Invalid characters are ignored until valid PHP code is found. Therefore, PHP and PHAR files can readily be combined with a number of file formats. 

Polyglots whose covert format is PHP or PHAR typically utilize an image format (JPEG, PNG, GIF, BMP) as their overt component, likely due to the prevalence and (possibly) lower level of scrutiny applied to images within web application file structures.  Since image files are commonly publicly accessible through file upload services, PHP and PHAR polyglots can serve as covert methods for staging and then executing malicious code on web servers. A web server's logs could merely show that a customer accessed a stored image when in reality they remotely executed malicious activity. Additionally, web servers that attempt to block malicious activity by preventing the upload of certain file formats are vulnerable to polyglots that masquerade as an approved format.

\subsubsection{JAR} 

The Java community created the Java Archive (JAR) to package a Java application, Java libraries, and other application resources in a single file. JAR files are an extension of the common Zip format. The contents of a Zip file are located by first scanning the end of the file for the \textit{central directory} which contains the relative offsets to the compressed files held within the archive, allowing another file to be prepended to an archive file without invalidating the data already contained in the archive.

Recently, threat actors created polyglots using JAR files. One possible reason is the discovery of the Windows vulnerability, CVE-2020-1464. This was a weakness in Windows Installer (MSI) files and is related to the manner by which their digital signature is validated within the Windows operating system. 

Normally, an MSI file is cryptographically signed by the developer, allowing an end-user to verify that the MSI not only came from the expected developer, but also has not been altered in transit. However CVE-2020-1464 allowed an attacker to append a malicious JAR file to the end of an MSI file without invalidating the signature of the MSI file, creating a polyglot with a covert format of JAR and an overt format of MSI. This vulnerability remained unpatched in the Windows operating system for at least two years. 

\subsubsection{Zip and RAR} 
The survey did not produce many instances of Zip and RAR polyglots. This may be due to the deletion of Zip and RAR polyglots once their contents have been extracted. In the attack chains observed with archive format polyglots, polyglot files with a covert format of Zip or RAR allowed covert transfer of the polyglot archive's malicious contents thanks, typically, to their image-based overt format.

Note, RAR files are not derived from Zip files; they are a distinct archive format. That said, RAR and Zip files both tolerate prepended data. Whereas Zip files are read from the bottom up, RAR files are read forward, skipping extraneous content until the RAR header is found. 
\subsubsection{JavaScript} 

The JavaScript language is a ubiquitous web technology used to build many web applications and is supported in all modern browsers. This provides a large attack surface for attackers. The survey discovered at least one instance of an attacker using a polyglot with a  JavaScript covert format and an image-based overt format to infiltrate advertisement networks.

Normally, reputable advertising companies restrict scripts in their advertisements to avoid sending end-users malicious code. However, this polyglot could bypass script detection without loss of functionality by posing as an advertising image. We were unable to get the sample for this attack. 


\begin{table}[h]
\caption {Malicious Polyglot Hashes}
\centering
\resizebox{\columnwidth}{!}{\begin{tabular}
{ |c|c|c|c| }
 \hline
 Malware Name & Formats & File Name & File Hash (SHA-256) \\
 \hline
 IcedId & CHM+HTA & pss10r.chm & 3d279aa8f56e468a014a916362540975958b9e9172d658eb57065a8a230632fa \\
 \hline
  Batloader & PE+HTA & AppResolver.dll & 1258fb78dd50f6c12c3181cc5c1362dc9d70ca46c5fd7e6af4880ee6d6d9e7a2 \\
 \hline
  Batloader & PE+HTA & AppResolver.dll & 588af958bc4365ecff4264a9fb75351eaee1ca9d0672c3040a77f979795219bd \\
 \hline
  Batloader & PE+HTA & AppVEntStreamingManager.dll & 3ec8b76ac735348db87bd0bf766554a2cb280f94d12dad8a159e917e00ab28f2 \\
 \hline
  ZLoader & PE+HTA & AppVSentinel.dll & 64a0a6ac17128ce7fd4dc34556bfe4736900121e5766557bceeac0cce99fbe21 \\
 \hline
  ZLoader & PE+HTA & AppResolver.dll & 89ccde97787a3eb0f9de38ab51c9f3278a3b18531b0fa468f08b55a133263b1c \\
 \hline
  ZLoader & PE+HTA & AppResolver.dll & 950ad539dfc8e16c07d24dbb37ae19daa0b2f32164ba0cb3c81fa7e689f274e1 \\
 \hline
  ZLoader& PE+HTA & AppResolver.dll & a187c9bb2a8bc29184bd18d6f515532d0f9b3f97b53f0ec6347b9982c4dff00f \\
 \hline
  ZLoader & PE+HTA & AppResolver.dll & c1a34057b31dd53e227a7001a7f0860e553b7efdb9ea2e9ec3b80221266b7d51 \\
 \hline
  ZLoader & PE+HTA & advapi32.dll & d1a1381c1f02abaa3449451136c1d1054ed72818348297113c135e8211173b3f \\
 \hline
  ZLoader & PE+HTA & AppResolver.dll & eb7354a95762565558d46753caf0c0d4dd09e1f358d564ae034b64446599e907 \\
 \hline
 Lazarus & BMP+HTA & imgFBE0.tmp & fe16b1dc30ee50ab126129c7fc0f2e6932083d4429241707d8046760c6b25042 \\
 \hline
 Lazarus & BMP+HTA & image003.zip - undetected & c9803b32365f4870d4ca833eb418eb845f16c4ec1628253a152667d935d9985b \\
 \hline
 Lazarus & BMP+HTA & image003.zip - undetected & a95a3fd25ab87c5010d42fe0131338b78187672dd6dc213af4253ef5db494591 \\
 \hline
 Lazarus & BMP+HTA & image003.zip & 888cfc87b44024c48eed794cc9d6dea9f6ae0cc3468dee940495e839a12ee0db \\
 \hline
  PHP Shell & JPEG+PHP & 63f4c7b002cc47.jpg & 4e26b08cce3fbd04fb9d954e1fa6a72d91f909015e7564aae9570aee26e8efd6 \\
 \hline
  PHP Shell & JPEG+PHP & simp.php.gif & 47102e200c35185654e74237a838e4c6b484cadd5a97d77aa7ad633b4f83ba62 \\
 \hline
  PHP Shell & JPEG+PHP & images.jpg & 0b5fd1d621affa41ebe811a39c085d62be489c55e26705b1db61accaa1dbcb6a \\
 \hline
  PHP Shell & JPEG+PHP & 001.swf.jpeg & 71f463e8d5c0f7ec6221a1cb9d5683766d5f7270ca80395bee5d0d00ec4ba0f3 \\
 \hline
  PHP Shell & JPEG+PHP & 20190225150235\_34013.php & 5f8e797b0f2b2efee4839841cc7b597f80b8b6f1558ec18b43a834e4bd540fdb \\
 \hline
  PHP Shell & JPEG+PHP & v1QR1M.gif & e028dc0e26b03a8a9cd5de11515f485dbaa57b721cb4ff4b1ffa115e64459eb9 \\
 \hline
  PHP Shell & GIF+PHP & Adipati.php & b660e691007a1fd8301f39782019a5f7bee6fd7dea18545e372a67014cee4c42 \\
 \hline
  PHP Shell & JPEG+PHP & Logo\_Coveright.jpg & ab85eb33605f3013989f4e8a9bfd5e89dd82d1f80231d4e4a2ceb82744bf287c \\
 \hline
  PHP Shell & JPEG+PHP & ce167d905d117823d780e188002b3120.jpg & 39588ed13465b15ec59ec35a885de028d0b6537cf6410c96402adfe1053694d6 \\
 \hline
  PHP Shell & PNG+PHP & in1.png & 57507a3db555182882c0c335b0b943ee2f977a1a9cf973be070fa9db6491cdf5 \\
 \hline
  SyncCrypt & JPEG+ZIP & 003\_JPG.arrival.jpg & c6565d22146045e52110fd0a13eba3b6b63fbf6583c444d7a5b4e3a368cc4b0d \\
 \hline
  DarkTrack RAT & PNG+RAR & darknet.jpg & ee0c0be30ba2875a2bc7813ae80814659ce35988fbd9d5232950ed7722b89a9a \\
 \hline
  JAR/MSI & MSI+JAR & 488adc.msi & dd71284ac6be9758a5046740168164ae76f743579e24929e0a840afd6f2d0d8e \\
 \hline
  Ratty & MSI+JAR & 29-05-2020.jar & 90f613caa131c663e32aabc31b5fccc99edcfa874110d51cd627531d3a67b16d \\
 \hline
  Ratty & MSI+JAR & 6afad7.msi & 04a3cad80470a085b6ef57a7e1007049a29863a94fe76f93be1f2a0c54da99d6 \\
 \hline
\end{tabular}}
\label{tab:survey_hashes}
\end{table}
\begin{table}[h]
\caption {Sources for Malicious Polyglot Usage in the Wild}
    \centering
    \resizebox{\columnwidth}{!}{
    \begin{tabular}{|p{10.0cm}|c|c|p{8.0cm}|}
    \hline
\textbf{Title} & \textbf{Publisher} &  \textbf{Date Published} & \textbf{URL} \\\hline
New Banking Trojan IcedID Discovered by IBM X-Force Research &  Security Intelligence & 13 November 2017 & \url{https://securityintelligence.com/new-banking-trojan-icedid-discovered-by-ibm-x-force-research/} \\\hline 

 More Than Meets the Eye: Exposing a Polyglot File That Delivers IcedID & Palo Alto Networks & 27 September 2022 & \url{https://unit42.paloaltonetworks.com/polyglot-file-icedid-payload/} \\\hline

Zoom For You — SEO Poisoning to Distribute BATLOADER and Atera Agent & Mandiant & 1 February 2022 & \url{https://www.mandiant.com/resources/blog/seo-poisoning-batloader-atera} \\\hline
 
Can You Trust a File’s Digital Signature? New Zloader Campaign exploits Microsoft’s Signature Verification putting users at risk & Check Point Research & 5 January 2022 & \url{https://research.checkpoint.com/2022/can-you-trust-a-files-digital-signature-new-zloader-campaign-exploits-microsofts-signature-verification-putting-users-at-risk/} \\\hline

BATLOADER: The Evasive Downloader Malware & VMWare & 14 November 2022 & \url{https://blogs.vmware.com/security/2022/11/batloader-the-evasive-downloader-malware.html} \\\hline
 Monitoring malware abusing CVE-2020-1599 & VirusTotal & 7 January 2022 & \url{https://blog.virustotal.com/2022/01/monitoring-malware-abusing-cve-2020-1599.html} \\\hline

Lazarus APT conceals malicious code within BMP image to drop its RAT & MalwareBytes & 19 April 2021 & \url{https://www.malwarebytes.com/blog/threat-intelligence/2021/04/lazarus-apt-conceals-malicious-code-within-bmp-file-to-drop-its-rat} \\\hline

Andariel evolves to target South Korea with ransomware & Kaspersky & 15 June 2021 & \url{https://securelist.com/andariel-evolves-to-target-south-korea-with-ransomware/102811/} \\\hline

LNK HTA Polyglot & Hatching & 12 November 2018 & \url{https://hatching.io/blog/lnk-hta-polyglot/} \\\hline

PHP WebShell Malware using Image Files & ASEC & 9 December 2020 & \url{https://asec.ahnlab.com/en/18861/} \\\hline

Hiding Webshell Backdoor Code in Image Files & Trustwave & 11 October 2013 & \url{https://www.trustwave.com/en-us/resources/blogs/spiderlabs-blog/hiding-webshell-backdoor-code-in-image-files/} \\\hline

Malware in Images: When You Can’t See "the Whole Picture" & Reversing Labs & 2 March 2021 & \url{https://blog.reversinglabs.com/blog/malware-in-images} \\\hline

Picture perfect: How JPG EXIF data hides malware &  Cisco & 24 July 2019 & \url{https://umbrella.cisco.com/blog/picture-perfect-how-jpg-exif-data-hides-malware} \\\hline

Lab: Remote code execution via polyglot web shell upload & PortSwigger & Unknown & \url{https://portswigger.net/web-security/file-upload/lab-file-upload-remote-code-execution-via-polyglot-web-shell-upload} \\\hline

Playing with GZIP: RCE in GLPI (CVE-2020-11060) & Almond & 14 May 2020 & \url{https://offsec.almond.consulting/playing-with-gzip-rce-in-glpi.html} \\\hline

It’s a PHP Unserialization Vulnerability Jim, but Not as We Know It & Blackhat & 9 August 2018 & \url{https://i.blackhat.com/us-18/Thu-August-9/us-18-Thomas-Its-A-PHP-Unserialization-Vulnerability-Jim-But-Not-As-We-Know-It.pdf} \\\hline

CVE-2022-41343 - RCE via Phar Deserialization & Tanto & 6 October 2022 & \url{https://tantosec.com/blog/cve-2022-41343/} \\\hline

Taiwan Heist: Lazarus Tools and Ransomware & BAE Systems & 16 October 2017 & \url{https://baesystemsai.blogspot.com/2017/10/taiwan-heist-lazarus-tools.html} \\\hline

SyncCrypt Ransomware Hides Inside JPG Files, Appends .KK Extension & Bleeping Computer & 16 August 2017 & \url{https://www.bleepingcomputer.com/news/security/synccrypt-ransomware-hides-inside-jpg-files-appends-kk-extension/} \\\hline

DarkTrack RAT – New Variant Thumbing a Ride in PNG Files & SECTRIO & 25 August 2020 & \url{https://www.subexsecure.com/pdf/malware-reports/August-2020/DarkTrack-Report.pdf} \\\hline

Distribution of malicious JAR appended to MSI files signed by third parties & VirusTotal & 15 January 2019 & \url{https://blog.virustotal.com/2019/01/distribution-of-malicious-jar-appended.html} \\\hline

Interesting tactic by Ratty \& Adwind for distribution of JAR appended to signed MSI – CVE-2020-1464 & Security-in-bits & 28 June 2020 & \url{https://www.securityinbits.com/malware-analysis/interesting-tactic-by-ratty-adwind-distribution-of-jar-appended-to-signed-msi/} \\\hline

Microsoft Put Off Fixing Zero Day for 2 Years & Krebs on Security & 17 August 2020 & \url{https://krebsonsecurity.com/2020/08/microsoft-put-off-fixing-zero-day-for-2-years/} \\\hline

GlueBall: The story of CVE-2020–1464 & Tal Be’ery & 16 August 2020 & \url{https://medium.com/@TalBeerySec/glueball-the-story-of-cve-2020-1464-50091a1f98bd} \\\hline

Uncovering and Disclosing a Signature Spoofing Vulnerability in Windows Installer: CVE-2021-26413 & Okta & 19 April 2021 & \url{https://sec.okta.com/articles/2021/04/uncovering-and-disclosing-signature-spoofing-vulnerability-windows} \\\hline
 
Hacking Group Using Polyglot Images to Hide Malvertising Attacks & Devcon & 24 February 2019 & \url{https://www.devcondetect.com/blog/2019/2/24/hacking-group-using-polyglot-images-to-hide-malvertsing-attacks} \\\hline

Bypassing Content Security Policy with a JS/GIF Polyglot & Ajin Abraham & 10 June 2015 & \url{https://ajinabraham.com/blog/bypassing-content-security-policy-with-a-jsgif-polyglot} \\\hline
 
WordPress Postie 1.9.40 Plugin - Persistent Cross-Site Scripting Exploit  & Vulners & 16 January 2020 & \url{https://vulners.com/zdt/1337DAY-ID-33819} \\\hline

CVE-2021-27190 – PEEL SHOPPING & Secuneus & 11 February 2021 & \url{https://www.secuneus.com/cve-2021-27190-peel-shopping-ecommerce-shopping-cart-stored-cross-site-scripting-vulnerability-in-address/} \\\hline

    \end{tabular}}
    \label{tab:survey_sources}
\end{table}

\end{document}